# Super-Resolution Coherent Diffractive Imaging via Titled-Incidence Multi-Rotation-Angle Fusion Ptychography

Youyang Zhou[1†], Weiren Shi[2†], Yun Xie[3], Bianli Zhao[4,5], Xinyu Luo[1], Mingjie Yao[6], Rui Zhang[4,5], Xin Tan[4,5], Kui Li[7], Hao Yang[8], Qi Liu[7], Yinggang Nan[1], Jie Bao[2], Yuping Zhang[2], Feng Shu[9], Shaopan Li[1*] & Xiaoshi Zhang[1,5,10*]

[1]School of Physics and Astronomy, Yunnan University, Yunnan, Kunming, 650500, China.
[2]School of Automation Engineering, University of Electronic Science and Technology of China, Chengdu, 611731, China.
[3]Academy for Engineering and Technology, Fudan University, Shanghai, 200433, China.
[4]School of Materials and Energy, Yunnan University, Yunnan, Kunming, 650500, China.
[5]South west United Graduate school, Yunnan, Kunming, 650092, China.
[6]School of Instrumentation and Optoelectronic Engineering, Beihang University, Beijing, 100191, China.
[7]School of Engineering, Yunnan University, Yunnan, Kunming, 650500, China.
[8]School of Chemical Science and Engineering, Yunnan University, Yunnan, Kunming, 650500, China.
[9]School of Electrical and Computer Engineering, The University of Sydney, Camperdown, NSW2006, Australia.
[10]Aerospace Information Research Institute, Chinese Academy of Sciences, Beijing, 100094, China
[†]These authors contributed equally.
*lishaopan5@gmail.com
*zhangxiaoshi@itc.ynu.edu.cn

**Abstract:** Coherent diffractive imaging (CDI) enables lensless imaging with experimental simplicity and a flexible field of view, yet its resolution is fundamentally constrained by the Abbe diffraction limit. To overcome this limitation, we introduce a novel Tilted-Incidence Multi-Rotation-Angle Fusion Ptychography technique. This approach leverages a tilted-incidence geometry to extend the collection angle beyond the Abbe limit, achieving up to a $\sqrt{2}$-fold resolution enhancement. By acquiring diffraction patterns at multiple sample rotation angles, we capture complementary spatial frequency information. A tilted-incidence multi-rotation-angle fusion ptychographic iterative engine (tmf-PIE) algorithm is then employed to integrate these datasets, enabling super-resolution image reconstruction. Additionally, this method mitigates the anisotropic resolution artifacts inherent to tilted CDI geometries. Our technique represents a novel advancement in super-resolution imaging, providing a novel alternative alongside established methods such as STED, SIM, and SMLM.

Coherent diffractive imaging (CDI) and its scanning variant, ptychography, are computational optical imaging techniques capable of reconstructing both the complex field of an object and the coherent illumination beam with near-wavelength resolution[1–8]. Compared to conventional CDI, ptychography significantly improves imaging versatility by allowing structured scanning, overcoming the limitation of requiring isolated samples. In ptychography, a sample is scanned with a coherent beam while ensuring a specific overlap between adjacent positions. The recorded diffraction patterns are then processed via iterative phase retrieval algorithms to reconstruct both the amplitude and phase of the sample.

Ptychography has become a general-purpose nano-imaging technique, particularly effective in the extreme ultraviolet (EUV) and soft X-ray (SXR) spectral regions. Unlike conventional imaging methods that rely on costly and low-quality EUV/SXR optics, ptychography benefits from an objective-lens-free design, offering a simplified imaging geometry. Consequently, research efforts have primarily focused on improving phase retrieval algorithms, making ptychography an indispensable tool at large-scale light source facilities such as synchrotrons and free-electron lasers (FEL)[9–12]. With the recent maturation of tabletop high-harmonic generation (HHG) sources, EUV/SXR ptychography[13–27] has evolved into a widely accessible nano-imaging technique, with applications spanning medicine, quantum information, nanoscience, advanced semiconductor technology and biosciences[28,29].

In recent years, the rapid advancement of Ptychographic Iterative Engine (PIE) algorithms[5,6], including mPIE[30], pc-PIE[31], zPIE[32], poly-CDI[33,34], mono-CDI[35,36], single-shot ptychography[37–42], ptychography with HDR[43], denoise[44,45] and machine learning[46–50], has solidified ptychography's role as a dominant nano-imaging technique. However, these algorithms predominantly assume normal-incidence imaging and rely on the paraxial approximation, which is unsuitable for tilted-incidence geometries.

Tilted-incidence imaging introduces asymmetric and distorted spatial frequency distributions at the detector. As the incidence angle increases, high-frequency components are preferentially collected above the sample plane, while spatial frequencies in other directions—especially below the sample plane—experience severe attenuation. This uneven distribution leads to anisotropic resolution degradation. Several strategies have been proposed to address these challenges. A tilted-plane correction (TPC) algorithm can mitigate spatial distortions before applying conventional phase retrieval model[15,51]. Additionally, rectangular detectors have been utilized to compensate for the preferential loss of spatial frequency components perpendicular to the sample plane[16,17,20,22]. However, these methods fail to recover the intrinsically lost diffractive signals that become evanescent waves under large-angle incidence conditions.

To address these fundamental limitations, we propose a novel tilted-incidence multi-rotation-angle fusion ptychography technique, coupled with a tilted-incidence multi-rotation-angle fusion iterative engine (tmf-PIE) algorithm, based on Ptylab[52]. In our method, a series of ptychographic diffraction images is acquired at multiple sample rotation angles in the tilted-incidence geometry. We first demonstrate that under optimal experimental conditions, this approach enables the acquisition of additional spatial frequency components beyond the conventional diffraction limit. Based on our soft threshold operation approach inspired by the soft thresholding procedure[53,54], we fuse diffraction datasets from at least two orthogonal sample rotation angles. This algorithm compensates for anisotropic high-frequency loss and restores a more uniform spatial frequency distribution. Notably, this technique does not require precise alignment between the sample rotation center and the illumination center, eliminating the need for frequency-domain image stitching.

We validate our method both theoretically and experimentally, demonstrate its ability to achieve super-resolution imaging and effectively mitigate resolution anisotropy caused by asymmetric spatial frequency sampling. To the best of our knowledge, this work represents the first proposal of a feasible solution for super-resolution CDI as well as the ptychographic algorithm for frequency domain fusion. We anticipate that tilted-incidence ptychography will emerge as a novel addition to the super-resolution imaging family, complementing established techniques such as single-molecule localization microscopy (SMLM)[55–59], structured illumination microscopy (SIM)[60–63], and stimulated emission depletion microscopy (STED)[64–66].

## Results

### Principle of Tilted-Incidence CDI and Its Latent Super-Resolution Potential.

In tilted-incidence coherent diffraction imaging (CDI), the sample plane and the detector plane are non-parallel. Under such geometry, conventional diffraction models—such as Fraunhofer diffraction, which assume parallel planes—are no longer valid. As a result, the recorded diffraction pattern cannot be directly employed in standard phase retrieval algorithms. To address this, a tilted-plane correction (TPC) must be applied to accurately interpret the acquired data. At large incidence angles, the system exhibits an intrinsic loss of spatial frequency components located below the sample plane. This leads to an asymmetric and distorted distribution of spatial frequencies in the diffraction pattern, particularly degrading resolution along the plane of incidence. As depicted in Fig. 1a, the region enclosed by the red and blue dashed lines indicates the inherent loss of diffracted information resulting from the large-angle geometry.

The far-field diffraction between two inclined planes is given by:

$$\tilde{\psi}(u,v) = \frac{1}{i\lambda} e^{i\left[kz + \frac{k}{2z}(u^2+v^2)\right]} \iint_{\Sigma} \psi(x,y) e^{-i\frac{2\pi}{\lambda z}(ux+vy)} dxdy \quad (1)$$

where $(x,y)$ denotes the sample spatial coordinates. The relationship between the spatial frequencies $(u, v)$ and the observation coordinates $(x',y')$ is described by the mapping[15]

$$\begin{cases} u = \frac{1}{\lambda}\left[\frac{x'}{r}\cos\theta + \left[\left(1 - \frac{x'^2 + y'^2}{R^2}\right)^{1/2} - 1\right]\sin\theta\right] \\ v = \frac{1}{\lambda} \cdot \frac{y'}{r} \end{cases}, \quad (2)$$

where $r = \sqrt{(x')^2 + (y')^2 + z_0^2}$ denotes the distance from the origin of the observation plane to the centre of the sample plane $(x, y, z)$, and $\theta \in (-\pi/2, 0]$ is the angle of incidence between the normal to the sample surface and the optical axis (cf. Fig. 1a). The results of the diffraction image obtained and after TPC are presented in Fig. 1b. It is evident that there is an asymmetry in the $u$- and $v$- directions within the collected frequency domain, which results in a decrease and asymmetry in resolution. The method of enhancing imaging resolution by expanding the acquisition range of diffraction information becomes ineffective.

Rearrange the above two expressions:

$$\begin{cases} r(\lambda u + \sin\theta) = x'\cos\theta + z_0\sin\theta \\ r(\lambda v) = y' \end{cases}, \quad (3)$$

By eliminating $x'$, a quadratic equation in $y'$ is formed

$$(\lambda u + \sin\theta)^2 + (\lambda v)^2 = \frac{1}{r^2}(x'\cos\theta + z_0\sin\theta)^2 + \left(\frac{y'}{r}\right)^2$$
$$= \frac{1}{r^2}\left[(x'^2 + z_0^2)\sin^2(\theta + \varphi) + y'^2\right] \leq 1/\lambda^2 \quad (4)$$

Where $\varphi = \arctan x'/z_0$.

Eq. (4) demonstrates that the presence of tilted incidence angle ($\theta$) results in an $\sin\theta/\lambda$ offset in the corresponding $u$-direction within the frequency domain space, $NA \to 1$. Additionally, the acquisition range in the frequency domain is represented as a circle centered at $(-\sin\theta/\lambda, 0)$ with a radius of $1/\lambda$. However we analyze the limiting conditions of Eq. (4) and find that the equality holds if and only if $\theta = 0, \varphi \to \pi/2$, i.e., normal incidence and $NA \to 1$, which implies that additional constraints/limitations exist. Delving deeper into this phenomenon, the upper limit of spatial resolution could be derived as

$$u^2 + v^2 = \left(\frac{x'}{r\lambda}\cos\theta + \frac{z_0}{r\lambda}\sin\theta - \frac{1}{\lambda}\sin\theta\right)^2 + \left(\frac{y'}{r\lambda}\right)^2$$
$$= \frac{1}{(r\lambda)^2}\left(\left[x'^2 + (r-z_0)^2\right]\cos^2(\theta+\beta) + y'^2\right), \quad (5)$$
$$\leq \frac{2}{\lambda^2}\left(1 - \sqrt{1 - NA^2}\right)$$

Where $\beta = \arctan\frac{r-z_0}{x'}$. The equal sign holds when $\theta = -\beta$, $\theta_0 = \arctan\frac{z_0-r}{x'}$. When $NA \to 1$, $\theta_0 = -\pi/4$, there exists point satisfying $u^2 + v^2 \to 2/\lambda^2 > 1/\lambda^2$, and $1/\lambda$ is the Abbe diffraction limit.

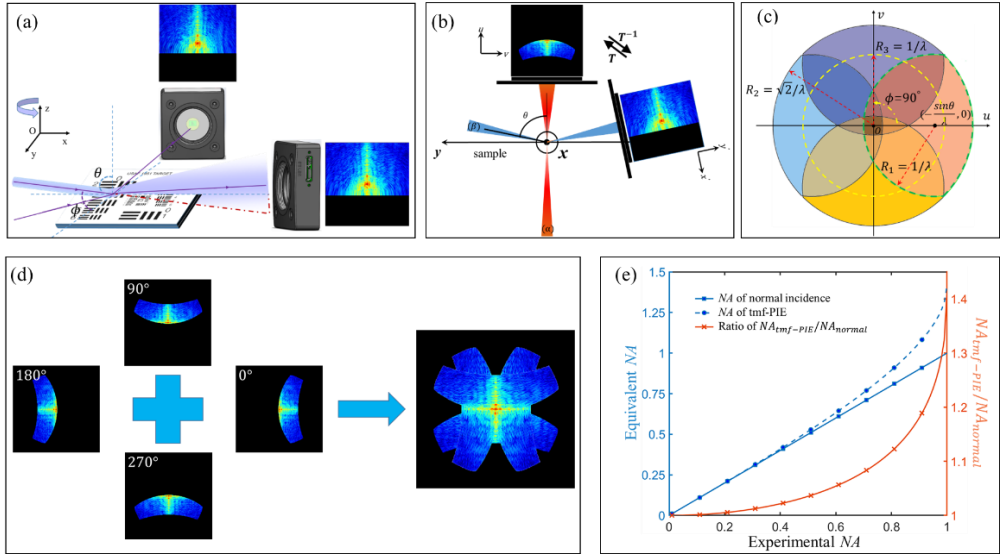

**Figure 1 | Principle and optical configuration of the tilted-incidence multi-rotation-angle fusion PIE (tmf-PIE) algorithm.** The tmf-PIE framework compensates for information loss due to evanescent waves and geometric constraints through a multi-angle fusion strategy, extending the spatial frequency domain beyond the conventional diffraction limit and enhancing resolution. (**a**) Schematic of the optical path for rotational fusion ptychographic imaging. $\theta$ is the angle of incidence between the normal to the sample surface and the optical axis angle and $\phi$ denotes the azimuthal angle between the projection of the incident beam in the $x$–$y$ plane and the negative $x$-axis. Illumination rotation and sample rotation are algorithmically equivalent. (**b**) Effect of large-angle tilted incidence on the recorded diffraction pattern and corresponding tilted-plane correction (TPC) to compensate for spatial frequency distortions. (**c**) Conceptual diagram showing the frequency domain shift caused by tilted incidence angle $\theta$ and the fusion of multi-angle diffraction data. The orange, purple, blue and yellow color filled region represents frequency coverage at $\theta = 0°$ and $\phi = 0°, 90°, 180°$ and $270°$ respectively with all its maximum frequency boundary at $NA = 1$. The yellow dashed line shows the maximum frequency domain boundary ($NA \to 1$) at normal ($\theta = 0°$) incidence. The tilted incidence geometry can surpass the Abbe theoretical limit by a factor of $\sqrt{2}$ (the black circle) in one direction such as the green dashed line and the black circle overlapped region,

while the multi-angle (0°, 90°, 180° and 270°) fusion allows the effective frequency boundary to surpass the theoretical limit in all directions. (**d**) Experimental multi-angle diffraction patterns captured under *NA* < 1 conditions, illustrating complementary spatial frequency content at different rotation angles. (**e**) Equivalent *NA* of imaging system and the ratio of the tilt-incidence enhanced *NA* v.s. normal imaging system *NA*. It shows that the higher the system *NA* the higher enhancement can be achieved using the tilt-incidence technique up to the limit of $\sqrt{2}$.

From Eq. (5), the theoretical resolution limit of our proposed algorithm can exceed the Abbe diffraction limit the yellow dashed circle as shown in Fig. 1c. In order to analyze the conditions for exceeding the Abbe diffraction limit, it is assumed that

$$u^2 + v^2 > \frac{1}{\lambda^2} \qquad (6)$$

Then the condition satisfied by Eq. (6) is:

$$\begin{cases} NA = \sin\alpha \geq \frac{\sqrt{3}}{2} \\ \theta \in \left(-\beta - \arccos\sqrt{K}, -\beta + \arccos\sqrt{K}\right) \cap \left(-\frac{\pi}{2}, 0\right] \end{cases}, \qquad (7)$$

Where $K = \frac{x'^2 + z_0^2}{x'^2 + (r-z_0)^2}$. In the case of *NA* greater than $\sqrt{3}/2$ and the incident angle $\theta$ satisfies Eq. (6), the titled-incidence CDI can capture the frequency-domain information beyond the the Abbe diffraction limit, and processed by tmf-PIE algorithm complete frequency domain information is a circle with a radius of $\sqrt{2}/\lambda$. And titled-incidence can always capture information farther away in the frequency domain space than the normal-incidence frequency domain. The detailed derivation process is shown in Supplement 1.

To intuitive understand the applicability of tmf-PIE, we simulate the variation trend of the frequency-domain information loss rate and the frequency-domain information distribution with respect to the incidence angle and *NA*. The calculation formula for the frequency-domain information loss rate is as follows:

$$SF_{Loss-Ratio} = 1 - \frac{S_{\theta,NA}}{S_{\theta,NA}|_{\theta=0, \ NA\to 1}} \qquad (8)$$

Here, $SF_{Loss-Ratio}$ represents the frequency-domain information loss rate and $S_{\theta,NA}$ represents represents the area projected into the frequency-domain space after TPC transformation.

The simulation results of the frequency domain information of the circular detector under different incidence angles and *NA* are shown in Fig. 2. From Fig. 2a, it can be seen that the frequency-domain information loss rate increases with increasing *NA* and incidence angle. Under the same *NA* condition, the loss rate varies approximately linearly with the incidence angle. As shown in Fig. 2b, when $\theta = 0°$, the frequency domain signal is circular and as the incident angle increases, the vertical width of the frequency domain signal narrows, while the horizontal width remains constant and gradually collapses into a crescent shape. The larger the angle, the larger the loss of frequency domain information in the ±*u* direction and ±*v* direction. Taking *NA* = 0.9 as example showed in Fig. 2c, which is one of essential prerequisites for achieving super-resolution, when $\theta$ around 20°, single view acquires the frequency information outside the area of frequency domain, which corresponds to $\theta = 0°$. And when $\theta$ between 20° and 80°, single view acquires the frequency information outside the area of frequency domain, which corresponds to Abbe diffraction limit.

**Titled-Incidence Multi-Rotation-Angle Fusion Ptychographic iterative engine (tmf-PIE).**
It follows from the properties of the Fourier transform that the resolution in the spatial domain $\Delta x$ is equal to the reciprocal of the width in the spatial-frequency domain *L*, i.e.,

$$L = \frac{1}{\Delta x} \tag{9}$$

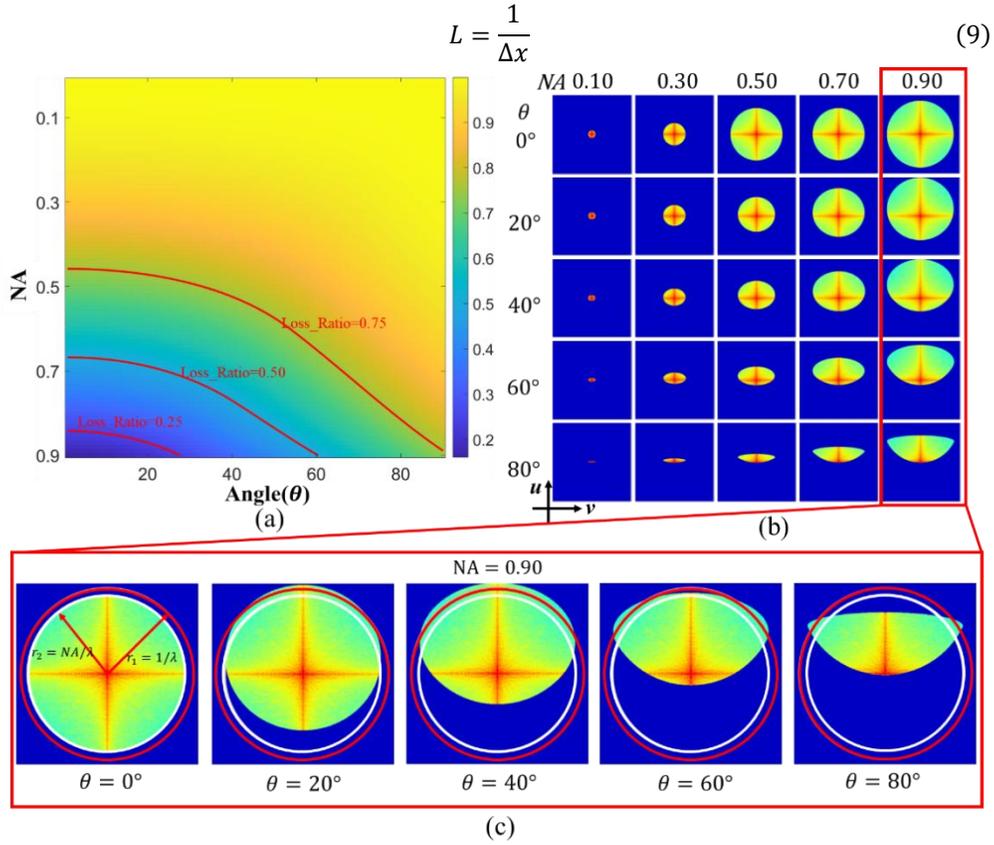

**Figure 2 | Simulation of frequency-domain information captured by a circular detector under varying incidence angles and numerical apertures (*NA*s).** The results demonstrate that tilted incidence can exceed the frequency range limitations of normal incidence, highlighting its potential for super-resolution imaging. (**a**) Loss rate of frequency-domain information as defined in Eq. (8) across different incidence angles and *NA*s. (**b**) Frequency-domain information captured at various incidence angles and *NA*s. (**c**) Magnified view of the frequency distribution at *NA* = 0.9 and $\phi = 90°$, under different incidence angles. The red circle indicates the Abbe diffraction limit, while the white circle represents the frequency range under normal incidence.

The absence of frequency signals asymmetrically alters the spectral and time-domain structure of the signal, often resulting in distortion in the time domain. Consequently, the resolution of the imaging result in a specific *r*-direction is typically determined by the narrower width of the positive and negative frequency information in that direction, i.e.

$$\Delta x_r = \frac{1}{2 min\{L_{+,r}, L_{-,r}\}} \tag{10}$$

It can be inferred that the resolution limit in the *x*-direction (corresponding to the *u*-direction) for titled-incidence coherent diffraction imaging, under an incident angle $\theta$, $NA \rightarrow 1$, is given by

$$\Delta x = \frac{\lambda}{2(1 - sin\theta)} \tag{11}$$

The resolution limit in the y-direction (corresponding to the *v*-direction) is

$$\Delta y = \frac{\lambda}{2\cos\theta} \quad (12)$$

According to equations (11) and (12), in titled-incidence CDI, as the incident angle increases, the resolutions in both the *x*-and *y*-directions reduce, and no longer the equal.

To address these issues, we propose tmf-PIE algorithm and an associated imaging technique. In our technique, we obtained a series of ptychographic diffraction patterns at different rotation angles, (e.g., $\phi = 0°, 90°, 180°, 270°$, as shown in Fig. 1c-d). We also show that the higher the system NA the higher enhancement can be achieved using the tilt-incidence technique up to the limit of $\sqrt{2}$ at $NA \to 1$ (Fig. 1e). These patterns were denoised and processed using the TPC algorithm. Subsequently, the diffraction patterns from every angle were labeled and integrated. Finally, the fused data were reconstructed using the tmf-PIE.
Based on equations (5), and (6), the frequency-domain range for titled incidence is further derived, which leads to

$$\begin{cases} \left(u + \frac{\sin\theta}{\lambda}\cos\phi\right)^2 + \left(v + \frac{\sin\theta}{\lambda}\sin\phi\right)^2 \leq 1/\lambda^2 \\ u^2 + v^2 \leq \frac{2}{\lambda^2}\left(1 - \sqrt{1-NA^2}\right) \end{cases}, \quad (13)$$

Here, $\theta$ represents the angle of incidence and $\phi$ represents the angle of rotation with respect to the initial incident light (defined as 0°). As shown in Eq. (11), the angle of incidence $\theta$ is kept constant and the incident light is rotated clockwise by $\phi$ so that the light is incident in the other direction. The projection on the frequency domain (*u, v*) corresponds to the intersection of two circular frequency domains. Under large-angle incidence conditions, the proportion of negative frequency signals decreases as $\sin\theta/\lambda$ approaches $1/\lambda$.

From Eq. (11), the radius of the large circle after rotation is obtained as:

$$R_2 = \min\left\{\frac{(1+\sin\theta)}{\lambda}, \frac{\sqrt{2}}{\lambda}\right\} (NA \to 1) \quad (14)$$

By substituting Eq. (12) into Eq. (8), the resolution of tmf-PIE imaging is obtained as:

$$\Delta r = \max\left\{\frac{\lambda}{2(1+\sin\theta)}, \frac{\lambda}{2\sqrt{2}}\right\} (NA \to 1) \quad (15)$$

By stitching frequency-domain information from different angles, the frequency-domain bandwidth of single-view imaging is extended, thereby improving the imaging resolution. Under ideal conditions ($NA > \sqrt{3}/2$), the imaging resolution exceeds the Abbe diffraction limit $\Delta r = \lambda/2NA$.

Based on the aforementioned content, we propose a tmf-PIE algorithm based on multi-rotation-angle fusion to enhance the resolution of coherent diffraction imaging under large-angle incidence conditions. Specifically, by keeping the incident angle constant and rotating the incident light source, frequency-domain information from different angles is acquired. The frequency-domain data from various perspectives are then combined complementary, thus obtaining a broader frequency-domain range.

The tmf-PIE algorithm extends the standard mPIE algorithm from two aspects. First, the object patch fetching and storing back are modified to take the current angle as an additional argument and execute using some kind of interpolation, e.g. bi-cubic interpolation. Second, when updating the wave field at the Fourier plane, a divergence is added to adopt two different update policies according to whether the current spatial frequency is acquired, called soft threshold operation.

Conventional Fourier plane projection algorithms assume the unknown regions to have zero values, which introduces conflicts within the multi-angle dataset. Specifically, the reconstructed object/probe frequency characteristics corresponding to different angle subsets become inconsistent, leading to poor reconstruction results or even failure of algorithm convergence.

To address this issue, we propose a novel frequency-domain information fusion method based on the soft threshold operation strategy—a well-known procedure for solving linear inverse problems[54]. In this approach, the unobserved frequency-domain information is set to not a number (NaN). During the projection operation in the frequency domain, normal projection is performed for the regions containing valid information. For the NaN regions, a discount factor is introduced to adjust the frequency-domain information. The process can be expressed as:

$$\widetilde{\Phi}_j^k(u,v) = \begin{cases} \sqrt{\dfrac{I_k}{\left|\Phi_j^k(u,v)\right|^2}}\Phi_j^k(u,v), & for\ \vec{r} \in detected\ zone \\ \sqrt{\alpha + (1-\alpha)\dfrac{\eta}{\left|\Phi_j^k(u,v)\right|^2 + \eta}}\Phi_j^k(u,v), & for\ \vec{r} \in NaN\ zone \end{cases}, \quad (16)$$

Where $I_k$ is the measured intensity, $\Phi_j^k(u,v)$ is the estimated value in the $j$th iteration, $\widetilde{\Phi}_j^k(u,v)$ is the updated value, $\alpha$ is the discount factor, which range is 0~1 and $\eta$ is the magnitude size determination threshold.

Eq. (16) not only retains the frequency-domain intersections during projection but also preserves the individual frequency-domain information from different perspectives, thereby forming a union of frequency domains.

**Numerical simulations of tmf-PIE in large angle of incidence.**
With the number of iterations set to 1000, the reconstruction results are shown in Fig.3a-b. Fig. 3a presents the amplitude distributions of the sample and probe reconstructed using tmf-PIE from a single angle, while Fig. 3b shows the amplitude distributions of the sample and probe reconstructed by fusing four rotation angles (0°, 90°, 180°, and 270°).

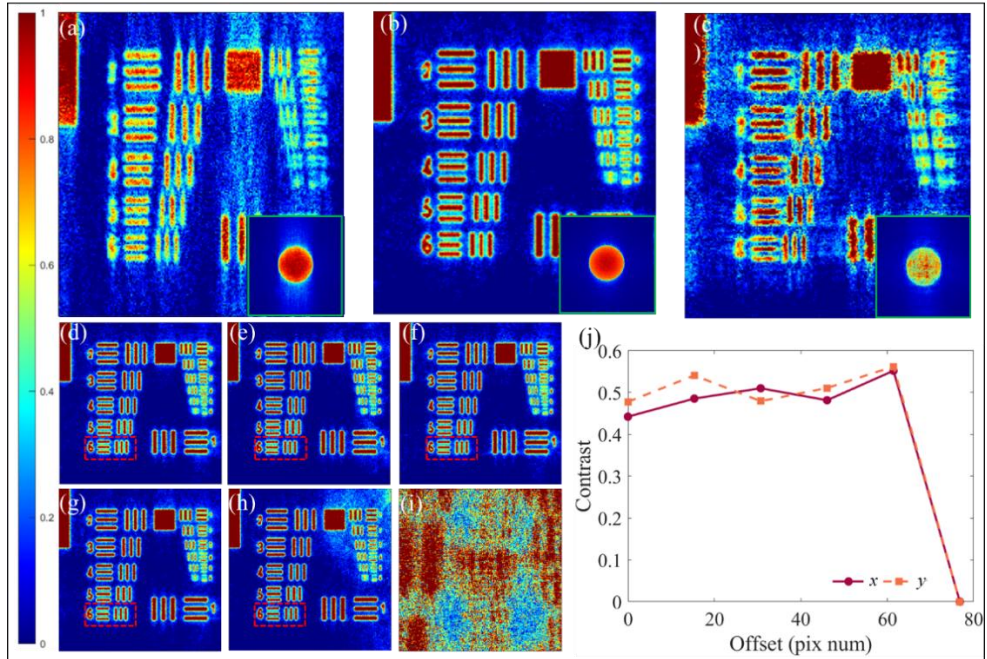

**Figure 3 | Simulation results demonstrating the performance of the tmf-PIE algorithm.** The reconstruction quality is substantially improved by fusing diffraction patterns acquired at four rotational orientations ($\phi$ = 0°, 90°, 180°, and 270°) and applying a soft thresholding operation in the frequency domain. This approach effectively suppresses the anisotropic

resolution artifacts characteristic of tilted CDI geometries. Furthermore, the algorithm exhibits strong robustness to spatial misalignment between the sample rotation axis and the illumination axis of the laser beam. (**a**) Reconstructed laser beam profile and sample amplitude using a single rotation angle. (**b**) Reconstructed laser beam and sample amplitude obtained by fusing data from four rotation angles with soft thresholding applied. (**c**) Reconstructed laser beam and sample amplitude from four rotation angles without applying the soft thresholding operation. (**d**)–(**i**) Reconstructed sample amplitude under increasing offset between the sample rotation center and the laser beam illumination axis: 0, 15, 30, 45, 60, and 75 pixels, respectively. (**j**) Quantitative evaluation of image quality in (**d**)–(**i**), expressed as average image contrast.

By comparing the single-angle and multi-angle reconstruction results, it is observed that the tmf-PIE algorithm successfully outputs their reconstructed results. However, the single-angle reconstruction exhibits significant noise, with severe trailing in grayscale variations between stripe regions and stripe spacing, leading to poor resolution performance. In contrast, the multi-angle reconstruction demonstrates clear improvements: on the left side (the cluster VI of the USAF 1951 resolution target), the numerical patterns are clearly restored, stripes are distinct, and the grayscale difference between stripe regions and stripe spacing is substantial. On the right side (the cluster VII of the USAF 1951 resolution target), both horizontal and vertical stripes in the first three groups (7-1, 7-2, 7-3) are clearly resolved. For the single-angle reconstruction, the grayscale difference between stripe regions and stripe spacing on the left side is minimal, the numerical patterns are blurred and indiscernible, and on the right side, only the vertical stripes of the first group (7-1) are barely distinguishable, while the horizontal stripes are difficult to resolve. Overall, the multi-angle reconstruction achieves a significant improvement in image quality compared to the single-angle reconstruction.

To further discuss the impact of soft threshold operation on the improvement of reconstructed image quality, the soft threshold operation component was set to zero during the simulation. The resulting image is shown in Fig. 3c. and it can be observed that without soft threshold operation, the background noise in the four-angle reconstructed image is relatively high. The stripes on the left are barely discernible, while on the right, only the vertical stripes of groups 7-1 can be resolved, whereas the horizontal stripes are entirely indistinguishable. Comparing Fig. 3a-c, it is evident that soft threshold operation and the multi-angle fusion strategy significantly enhances the resolution of reconstructed images. Without soft threshold operation, even using the multi-angle reconstruction method, the resolution improvement remains limited, though still better than that achieved with the single-angle method.

In the numerical simulation, it was assumed that the illumination center and the rotation center are in perfect coincidence effortlessly, which is extremely difficult to achieve in practical experiments. The offset between the illumination center and the rotation center causes position errors when transforming position datasets from different angles into a unified coordinate system. In order to investigate the impact of these errors on our algorithm, we introduce an offset between different dataset to investigate the effect of different offset magnitudes on reconstruction quality. The results are shown in Fig. 3d-i. From the figures, it can be observed that as the offset increases, the noise in the reconstructed image of the seventh cluster of the USAF 1951 resolution target gradually intensifies, though the stripes remain distinguishable. When the offset becomes excessively large (up to 75 pixels), the reconstruction fails completely. Fig 3j shows the variation curve of the reconstructed stripe contrast for the sixth group of the sixth cluster as a function of offset. The curve indicates that when the offset is less than 60 pixels, the changes in stripe contrast are minimal, whereas when the offset exceeds 60 pixels, the stripe contrast rapidly deteriorates, ultimately rendering reconstruction impossible.

It is noteworthy that tmf-PIE takes effect in the overlap field of view (FOV) region of different rotation angle. And generally, the best reconstruction in ptychography is in the center region. Thus the closer the alignment between the illumination center and the rotation center, the more pronounced the algorithm's effectiveness becomes.

**Experimental example with the USAF-1951 resolution target and a chip.**

In order to further verify the accuracy and effectiveness of the proposed tmf-PIE algorithm, as illustrated in Fig. 4, the large-angle titled-incidence ptychographic imaging optical setup was constructed for experimental validation. Fig. 5a-c illustrate the results of the reflectance imaging recovery reconstruction for the USAF-1951 resolution target experiments conducted at $\phi = 0°$, 90°, and 0°+90°. The stripes from the first three element s of the Cluster VII—7-1 (spaced at 3.91 µm), 7-2 (3.48 µm), and 7-3 (3.10 µm)—are clearly visible. As illustrated in Fig. 5, the two peaks (i.e., the three tick marks) can be clearly identified, regardless of whether they are oriented in the vertical or horizontal direction.

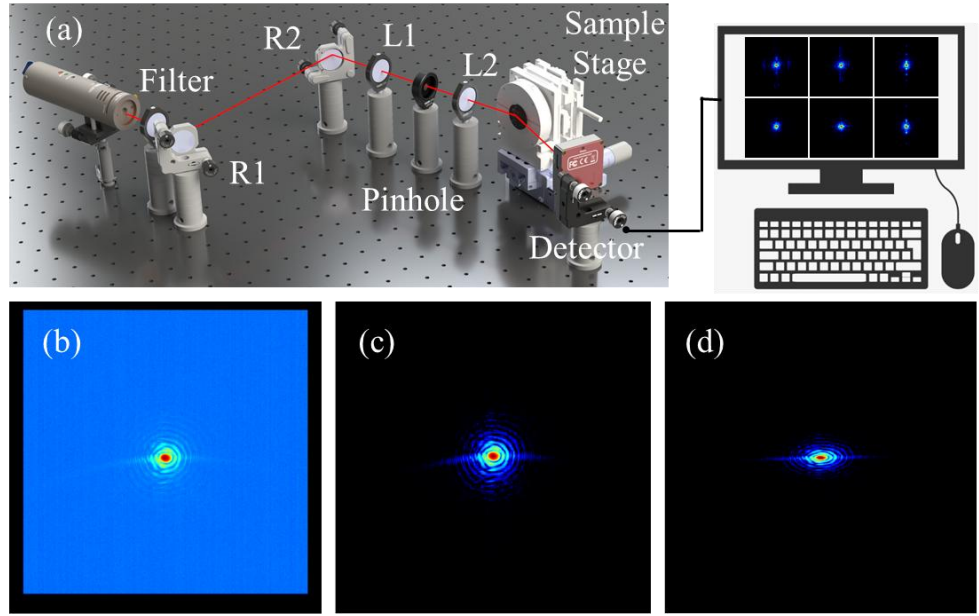

**Figure 4 | Experimental setup for the Tilted-Incidence Multi-Rotation-Angle Fusion Ptychography technique and corresponding raw image data processing workflow.** (**a**) Schematic of the experimental setup. (**b**) Raw diffraction image acquired during the experiment. (**c**) Denoised diffraction image following the first-stage denoising process. (**d**) Diffraction image after applying Tilted Plane Correction (TPC) in the second processing step.

To further discuss the resolution of the sample amplitudes recovered by the tmf-PIE algorithm, the grayscale distributions of the transverse and longitudinal stripes of the seventh cluster of 7-3 stripes (spaced 3.10 µm apart) are plotted, as shown in Fig. 5d-e. As illustrated in the figure, the vertical resolution exhibits the greatest difference in grayscale values between the peaks and troughs of the reconstructed image at a rotation angle of 90°. This angle also demonstrates the most pronounced contrast among the stripes. Conversely, the contrast of the stripes in the reconstructed image at a rotation angle of 0° is the lowest.

In addition, image reconstruction was performed on a defective chip sample. Fig. 6a presents a microscope image of a specific region of the chip, where the black spots indicate defects caused by ablation. A detailed analysis of the reconstruction results at $\phi = 0°$, 90°, and 0° + 90° rotations reveals that the 0° + 90° reconstruction provides the clearest image, successfully capturing the contours and variations of complex patterns, such as the V-shaped pattern and the left 'L' pattern. In contrast, the 0° reconstruction suffers from the loss of horizontal frequency domain information, resulting in blurring in the upper portion of the V-shaped pattern and horizontal trailing in the left 'L' pattern. Similarly, the 90° reconstruction exhibits loss of vertical frequency domain information, causing blurring in the lower portion of the V-shaped pattern and vertical trailing in the left 'L' pattern.

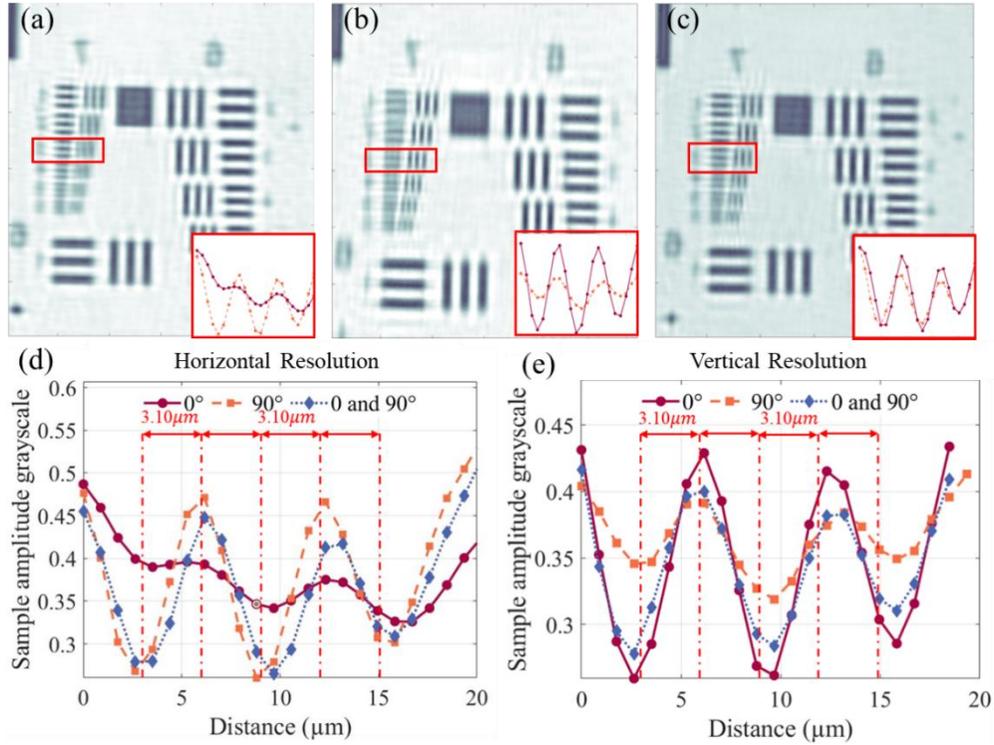

**Figure 5 | Experimental results: Amplitude reconstruction of USAF1951 resolution targets.** The results demonstrate the algorithm's capability to suppress anisotropic resolution artifacts intrinsic to tilted CDI geometries. (**a**) Reconstructed image of Group VII of the USAF1951 target at sample rotation angle φ = 0°. (**b**) Reconstructed image of Group VII at φ =90°. (**c**) Reconstructed image of Group VII obtained by fusing multiple sample rotation angles at φ = 0°and 90°. (**d**) and (**e**) Intensity line profiles extracted from the horizontal and vertical bars of the third element in Group VII, corresponding to (**a**)–(**c**), respectively.

For chip defects, all three reconstruction approaches can identify the defect locations. However, their ability to depict the shapes of defects varies. All three methods accurately reconstruct the smaller defects in the upper-left and central regions, but struggle to delineate the shapes of the larger defects in the lower-left and lower-right regions. This limitation arises because the deeper ablation of the larger defects causes their frequency domain projections to deviate from those of the surrounding areas, resulting in discrepancies between the reconstructed images and the microscope observations.

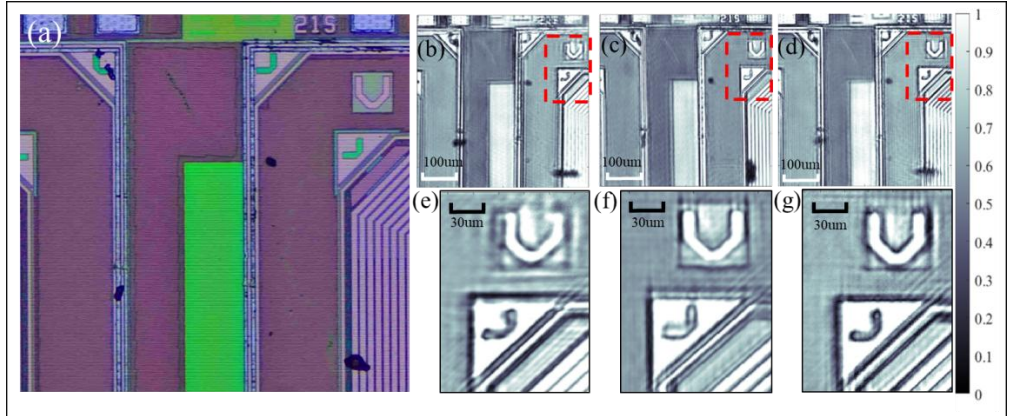

**Figure 6 | Reconstructed images of a computer chip sample.** Comparison between single-angle and multi-angle reconstructions demonstrates that the proposed multi-angle fusion approach significantly enhances resolution and achieves isotropic resolution across all directions. (**a**) Optical microscope image of the chip sample. (**b**) and (**c**) Reconstructed images obtained using the experimental setup at sample rotation angles of $\phi = 0°$ and $\phi = 90°$, respectively. (**d**) Reconstructed image obtained by fusing data from $\phi = 0°$ and $\phi = 90°$ using the tmf-PIE algorithm. (**e**)–(**g**) Zoomed-in views of the regions highlighted by the red dashed boxes in (**b**)–(**d**), respectively.

## Discussion

In this work, we investigate diffraction patterns propagating between tilted planes and analyze the anisotropic loss of spatial frequency information caused by interpolation blindness under large-angle incidence, using the Tilted Plane CDI (TPC) technique based on the diffractive "Edward" sphere model. Building on these studies, we introduce a novel tilted-incidence ptychographic imaging method and develop the multi-rotation-angle fusion phase iterative engine (tmf-PIE) algorithm. This approach overcomes the limitations of conventional collection-angle-constrained coherent diffraction imaging (CDI), achieving spatial resolution significantly beyond the classical diffraction limit.

We demonstrate that our tilted-incidence imaging technique, together with the tmf-PIE algorithm, enables super-resolution imaging up to a [X]-fold improvement beyond the Abbe limit, while also addressing the inherent resolution anisotropy associated with large-angle-incidence geometries.

To validate our approach, we constructed an experimental setup for tilted-incidence CDI using visible laser light and conducted imaging experiments on a USAF1951 resolution target and a computer chip sample. The results confirm the capability of our method to achieve enhanced and isotropic resolution under large-angle incidence, validating the effectiveness of the tmf-PIE algorithm in practical applications.

Looking forward, several strategies can further push the resolution beyond the Abbe limit:

(a) increasing the system's numerical aperture (NA) beyond 0.5 to realize the tilt-incidence-enhanced NA as predicted by our theory;

(b) employing coherent structured illumination to emulate tilt-incidence effects;

(c) exploiting nonlinear sample responses, such as second- or third-harmonic generation (SHG/THG);

or (d) combining these strategies to achieve a substantial resolution boost.

We anticipate that CDI, empowered by our proposed methodology, will emerge as a compelling alternative to established super-resolution techniques such as STED, SIM, and SMLM, offering advantages in both resolution and scalability.

Our findings underscore the potential of tilted-incidence configurations to fundamentally extend the spatial resolution limit in lensless imaging. This approach is particularly promising for applications in quantum material characterization and single-molecule bioimaging, where non-destructive, high-resolution visualization of delicate nanostructures and dynamic processes is essential. Furthermore, the inherent scalability of our method—compatible with photon energies from visible light to EUV and X-ray regimes—lays the foundation for a versatile, universal framework adaptable to diverse experimental requirements. This breakthrough redefines the capabilities of lensless imaging and opens new avenues for real-time nanoscale observation across disciplines, including physics, chemistry, quantum information, and the life sciences.

## Methods
### Setting of numerical simulation.
The simulation parameters are as follows: the incidence angle is $\theta = 78°$, the illuminated probe wavelength is $\lambda = 445$ nm, the spot diameter is 315 μm, the samples is a standard USAF 1951 resolution negative target, the diffraction distance is $z_0 = 3.5$ mm, camera imaging dimensions are $5.55\text{mm} \times 5.55\text{mm}$, the NA is about 0.63, the pixel size is $28.06\mu\text{m} \times 28.06\mu\text{m}$ as well as the scanning path is the Fermat spiral. The simulation is performed on a laptop, which processor is AMD Ryzen 7 5800H 3.20 GHz, installed RAM is 32.0GB, graphics is NVIDIA GeForce RTX 3060 Laptop.

The sample amplitude distribution is a standard USAF 1951 resolution negative target, represented as a $2048 \times 2048$ pixel matrix. To correspond with the frequency domain plane ($u, v$) and facilitate the multiplication of the illumination probe with the local sample matrix, the illumination probe is set as a $1024 \times 1024$ square matrix. The ePIE algorithm is employed for layered scanning, with the scanning path following a Fermat spiral and 70% overlap. A single scan yields 144 diffraction patterns at different scanning positions. The sample is rotated by 90°, 180° and 270°, and the original scanning coordinates are used for subsequent scans, resulting in a total of $144 \times 4 = 576$ diffraction patterns.

### Experiment setup.
The laser wavelength is 635nm and power is 5mW, which is generated by LM15-635 from LBTEK. L1 is a convex lens with a focal length of 120 mm while L2 is 60 mm. The pinhole is positioned at the focal point of L1, which diameter is 100 μm. In this configuration, the pinhole, L2, and the sample form a 4F system. In our experiment, the angle of incidence of the laser is 60°, The distance from the sample to the Detector (Daheng Optics, HD-G230M-GigE, 1/1.2") is 15.6mm. In the ptychography imaging with tilted-incidence geometry shown in Fig. 4, the stage was scanned using a positive resolution target, the USAF-1951 (LBTEK, RB-N). The grid was shifted according to the Fermat solenoids, with a gap of approximately 20 um, ensuring an overlap rate of over 70%.

Under this experiment setup-$NA \approx 0.22$ and incidence angle is 60°, the significant disparity in resolution along the *x*- and *y*- directions provides an appropriate platform for experimentally validating the effectiveness of the proposed algorithm. Furthermore, given that the asymmetry between the negative and positive frequency domains under these experimental conditions is not severe, we strategically selected datasets with rotation angles of 0° and 90° for the tmf-PIE trials after comprehensively balancing resolution requirements with computational efficiency.

**Reconstruction algorithm execution flow.**
Normally we start with a random initial guess at the object plane and proceed as follows for the $j$-th iteration,

(a) Update the probe and object

$$\begin{cases} O_j(\mathbf{r}-R_k) = O_j(\mathbf{r}-R_k) + \beta \cdot F\left(\dfrac{P_{j-1}(\mathbf{r})}{\left|P_{j-1}(\mathbf{r})\right|^2_{max}}(\widetilde{\Psi}^{k,l}_{j-1} - \Psi^{k,l}_{j-1}), l\right) \\ P_j(\mathbf{r}) = P_{j-1}(\mathbf{r}) + \dfrac{\beta F^{-1}(O_{j-1}(\mathbf{r}-R_k), l)}{\left|F^{-1}(O_{j-1}(\mathbf{r}-R_k), l)\right|^2_{max}}(\widetilde{\Psi}^{k,l}_{j-1} - \Psi^{k,l}_{j-1}) \end{cases}, \quad (17)$$

Where $P_j$ and $O_j$ is the estimate probe and object and $l$ corresponds to the rotation angle. $\Psi^k_{j-1}$ is the estimate exit wave and $\widetilde{\Psi}^k_{j-1}$ is the estimate after revising by the measured data in the $(j-1)$-th iteration. The constant $\beta$ is used to control the rate. The function $F$ rotates the matrix to $0°$ rotation angle according to the label $l$ via interpolation.

(b) Propagate the exit wave to the Fourier plane using the propagation model,

$$\Phi^k_j = P_z(P_j(\mathbf{r})O_j(\mathbf{r}-R_k)) \quad (18)$$

where $\Phi^k_j$ represents the Fourier plane wave field.

(c) According to the angular label of measured intensity $I^l_k$ refine the Fourier plane wave field with the soft threshold operation mentioned in Eq. (16)

$$\widetilde{\Phi}^{k,l}_j = softprocess(\Phi^k_j, I^l_k) \quad (19)$$

(d) Propagate the refined detector plane wave field back to the object plane

$$\widetilde{\Psi}^{k,l}_j = P_z^{-1}(\widetilde{\Phi}^{k,l}_j) \quad (20)$$

**Data availability.** All the relevant data are available from the authors on request.

**Acknowledgements**
This research was supported by grants from the Beijing Municipal Science & Technology Commission, Administrative Commission of Zhongguancun Science Park (No. Z241100004724002); the Yunnan Provincial Major Scientific and Technical transfer founding (No. FWCY-ZD2024002); the National Key Research and Development Program of China (14th Five-Year Plan, No. 2021YFB3602600); the Scientific Research Fund Project of Yunnan Education Department (No. 2025Y0068); and the Scientific Research and Innovation Project for Postgraduate Students at Yunnan University. The authors gratefully acknowledge the crucial financial and technical contributions of these organizations to the successful completion of this work.

**Author contributions**
Y.Z. and W.S. conceived the research idea together. W.S. implemented the algorithm through coded solutions and conceived the numerical simulation; experiments were designed by Y.Z.; Y.X. analyzed the data; the experiment was carried out by Y.Z. and Y.X.; B.Z. contributed to model design and sample customization; M.Y., R.Z., and X.T. maintained the experimental equipment; K.L. Y.N. and Q.L. offered academic guidance and statistical analysis support; X.L.

assisted with equipment model construction and drawing; H.Y. provided sample support; F.S., J.B. and Y.Z. managed project coordination; S.L. and X.Z. supervised the study, guided the analysis, and edited the manuscript; Y.Z., S.L. and X.Z. wrote the manuscript and Supplementary Materials with contributions from all authors.

## Additional information

**Supplementary Information**

**Competing financial interests:** The authors declare no competing financial interests.